\documentclass[12pt]{article}

\usepackage{newtxtext,newtxmath}

\usepackage{graphicx}

\usepackage[letterpaper,margin=1in]{geometry}

\renewenvironment{abstract}
	{\quotation}
	{\endquotation}

\date{}

\makeatletter
\renewcommand{\fnum@figure}{\textbf{Figure \thefigure}}
\renewcommand{\fnum@table}{\textbf{Table \thetable}}
\makeatother

\usepackage{scicite}

\usepackage{url}

\def\scititle{
Observation of Half-Quantum Vorticity in an Iron-based Superconductor 
}
\title{\bfseries \boldmath \scititle}

\author{
	Mohammad~Javadi~Balakan$^{1\ast}$,
	Genda~Gu$^{2}$,
        Qiang~Li$^{2,3}$,\and
        Kenji~Watanabe$^{4}$,
        Takashi~Taniguchi$^{4}$,
        Ji~Ung~Lee$^{1\dagger}$\and
        \small$^{1}$College of Nanotechnology, Science, and Engineering, State University of New York, Albany, NY 12203, USA.\and
        \small$^{2}$Condensed Matter Physics and Materials Science Division, Brookhaven National Laboratory, Upton, NY 11973, USA.\and
        \small$^{3}$Department of Physics and Astronomy, Stony Brook University; Stony Brook, New York 11794, USA.\and
        \small$^{4}$National Institute for Materials Science, 1-1 Namiki, Tsukuba 305-0044, Japan.\and       
	\small$^\ast$Corresponding author. Email: mjavadibalakan@albany.edu\and
	\small$^\dagger$jlee1@albany.edu.
}

\begin{document} 

\maketitle

\begin{abstract} \bfseries \boldmath
Half-quantum vortices---topological excitations carrying half the superconducting flux quantum---are predicted to emerge in spin-triplet superconductors, where the spin component of order parameter enables fractional flux quantization. We present direct transport signatures of half-quantum vorticity in single-crystal Fe(Te,Se), an iron-based superconductor with helical Dirac surface states. Using mesoscopic superconducting ring devices, we observe half-integer quantum oscillations arising from the splitting of quantized fluxoid states, further supported by distinct signatures of trapped half-integer fluxoids. The quantum oscillations are modulated by a background symmetry, set by the relative orientation of the DC bias current and perpendicular magnetic field, consistent with strong spin-orbit coupling. These findings demonstrate the long-sought half-quantum vorticity in spin-polarized superconductors and open new directions toward non-Abelian excitations in quantum materials and scalable topological qubits.
\end{abstract}

\noindent
The macroscopic wavefunction of a superconductor necessarily maps onto itself upon circulating around a ring, resulting in a quantized magnetic flux of $\Phi_0=h/2e$, where $h$ is Planck constant and $e$ is the elementary charge. A hallmark of triplet superconductivity is that the spin degree of freedom can participate in the phase winding, allowing for an exotic case in which the enclosed flux equals half of the superconducting flux quantum (Fig.~\ref{fig_sem}A and \ref{fig_sem}B). Half-quantum vorticity (HQV)---arising from a $\pi$ spin disclination in the $\mathbf{d}$-vector of a triplet superconductor---is a topologically stable singularity~\cite{volovik_1976, volovik_1999}. The core of a half-quantum vortex localizes a Majorana zero mode with non-Abelian statistics~\cite{ivanov_2001, readgreen_2000}, which serves as a key ingredient for topologically protected, fault-tolerant quantum computation~\cite{kitaev_2002,nayak_2008,alicea_2012,beenakker_2013,yazdani_2023}. Despite their fundamental and technological significance, half-quantum vortices have remained elusive in phase-sensitive transport measurements, largely due to challenges in thermodynamic stability and the subtle nature of their experimental manifestations~\cite{sigrist_1991,chung_2007}. 
While Cantilever magnetometry on strontium ruthenate revealed half-height magnetization steps in the presence of an in-plane magnetic field~\cite{jang_2011}, transport studies on doubly-connected superconducting rings did not exhibit robust half-quantum oscillations~\cite{yasui_2017, yasui_2020,cai_2022}. The formation of a single HQV in the bulk is thermodynamically unfavorable because of the logarithmically divergent energy associated with the unscreened spin current~\cite{sigrist_1991}. However, a pair of HQVs with opposite windings is theoretically predicted to eliminate this divergence~\cite{chung_2007}. 

In this work, we present compelling evidence of half-quantum oscillations in multiply-connected superconducting devices based on single-crystal Fe(Te$_{0.55}$,Se$_{0.45}$), achieved through a meticulous device fabrication process and precision transport measurements. As an iron-based superconductor~\cite{fernandes_2022}, Fe(Te,Se) provides a promising single-material platform for topological superconductivity evidenced by helical Dirac surface state~\cite{wang_2015,zhang_2018,zhang_2019,zaki_2021,li_2021} and signatures of Majorana bound states in the vortex core~\cite{wang_2018,machida_2019,kong_2019,zhu_2020, wang_2020}. Our direct measurement of half-quantum vorticity, accompanied by evidence of trapped half-quantum fluxoid, unambiguously reveals the topological nature of vortices in Fe(Te,Se). Furthermore, magnetoresistance measurements exhibit the splitting of individual full-quantum fluxoid states into pairs of HQVs, with clear evidence of HQV formation at zero bias. We find that the quantum oscillations exhibit a symmetry-dependent modulation governed by the relative orientation of the DC bias current and the perpendicular magnetic field. Finally, we identify two distinct sets of quantum oscillations, suggesting the presence of multiple Fermi surfaces with slightly different transition temperatures.

\subsection*{Device structure and measurements}
\begin{figure}
    \centering
    \includegraphics[width=.8\textwidth]{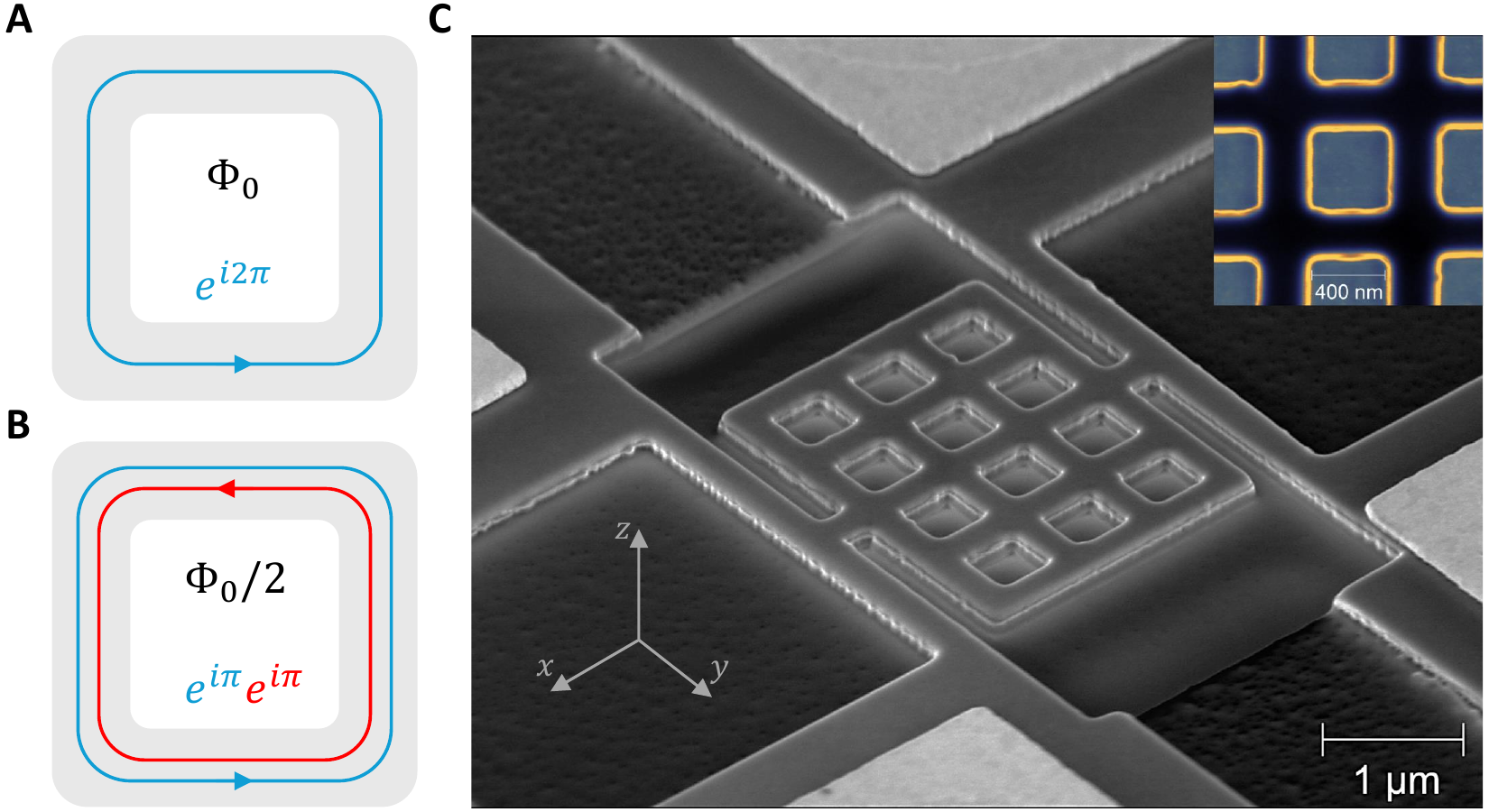}
    \caption{\textbf{Phase winding and device structure.} Phase winding of the superconducting order parameter for (\textbf{A}) a conventional superconductor, where only the orbital component contributes to the phase winding, and (\textbf{B}) a spin-triplet superconductor, where both orbital and spin degrees of freedom contribute. (\textbf{C}) An scanning-electron micrograph of a multiply-connected superconducting device. Inset shows a top-view image (false-color) of the rings.}
    \label{fig_sem}
\end{figure}

Fig.~\ref{fig_sem}C presents a micrograph of a multiply-connected superconducting device. The device consists of identical interconnected rings spanning an area of approximately 5~$\mu m^2$ to minimize spatial stoichiometric inhomogeneity~\cite{li_2021}. The Fe(Te,Se) single-crystal is fully covered by a thin hBN layer to prevent degradation~\cite{methods}. The device architecture shown in Fig.~\ref{fig_sem}C mitigates the impact of structural asymmetry by averaging over an array of rings, while simultaneously enabling the emergence of several HQVs, which is predicted to stabilize half-quantum oscillations through pairing of half-quantum vortices with opposite sense windings~\cite{chung_2007}. Fluxoid quantization in a multiply-connected superconductor gives rise to oscillatory magnetoresistance near the transition temperature ($T_c$), with a periodicity set by the quantum of magnetic flux~\cite{littlepark_1962, littlepark_1964, moshchalkov_1995}. Recognizing the close correspondence between fluxoid states in an annular geometry and vorticity of vortices in the bulk, we will use full-quantum vorticity (FQV) and half-quantum vorticity (HQV) to denote integer and half-integer fluxoid states, respectively.

An HQV is expected to manifest as a small feature in the magnetoresistance oscillations with a significantly low amplitude~\cite{cai_2022}. This feature, however, arises from a sharp change in the superfluid velocity of the superconductor. In this context, a direct measurement of variations in the superfluid velocity (or free energy) offers a feasible approach to the detection of HQVs. These variations can be measured experimentally using an AC lock-in technique, where the second-harmonic response to an AC excitation carries nonlinear transport information by capturing the gradient of the fundamental harmonic with respect to an independent variable. Generally, both Little-Park~\cite{littlepark_1962,littlepark_1964} and vortex-crossing~\cite{sochnikov_2010,berdiyorov_2012} effects contribute to the oscillatory behavior of magnetoresistance in a multiply-connected superconductor. The second-harmonic response ($V_{2\omega}$), is proportional to the superfluid velocity in the former case, and to the changes in free energy of the superconductor in the latter~\cite{methods}.

Electrical measurements were performed using a four-terminal method to eliminate contact resistance. A constant AC excitation of 1 $\mu$A, superimposed on a variable DC current, is used to probe the phase space~\cite{methods}. In the following, we represent the measurements on Device 1, which consist of rings with an inner side length of 406 nm and a wall width of 118 nm, corresponding to a superconducting flux quantum of $\sim$ 70 Gauss. Device 1 exhibits a normal-state resistance of $\sim$ 600 $\Omega$, with a transition onset at 14 K and a fully developed superconducting-state at $T_c=8$ K (Fig.~\ref{fig_mr7K}A). Prior to the magnetoresistance experiments, the device is cooled down from room temperature in an in-plane ($+\hat{y}$) magnetic field of 1 Tesla. This initial field-cooling establishes the symmetry for the emergence of half-integer fluxoid states.

\subsection*{Half-Quantum Oscillations}
\begin{figure}
    \centering
    \includegraphics[width=\textwidth]{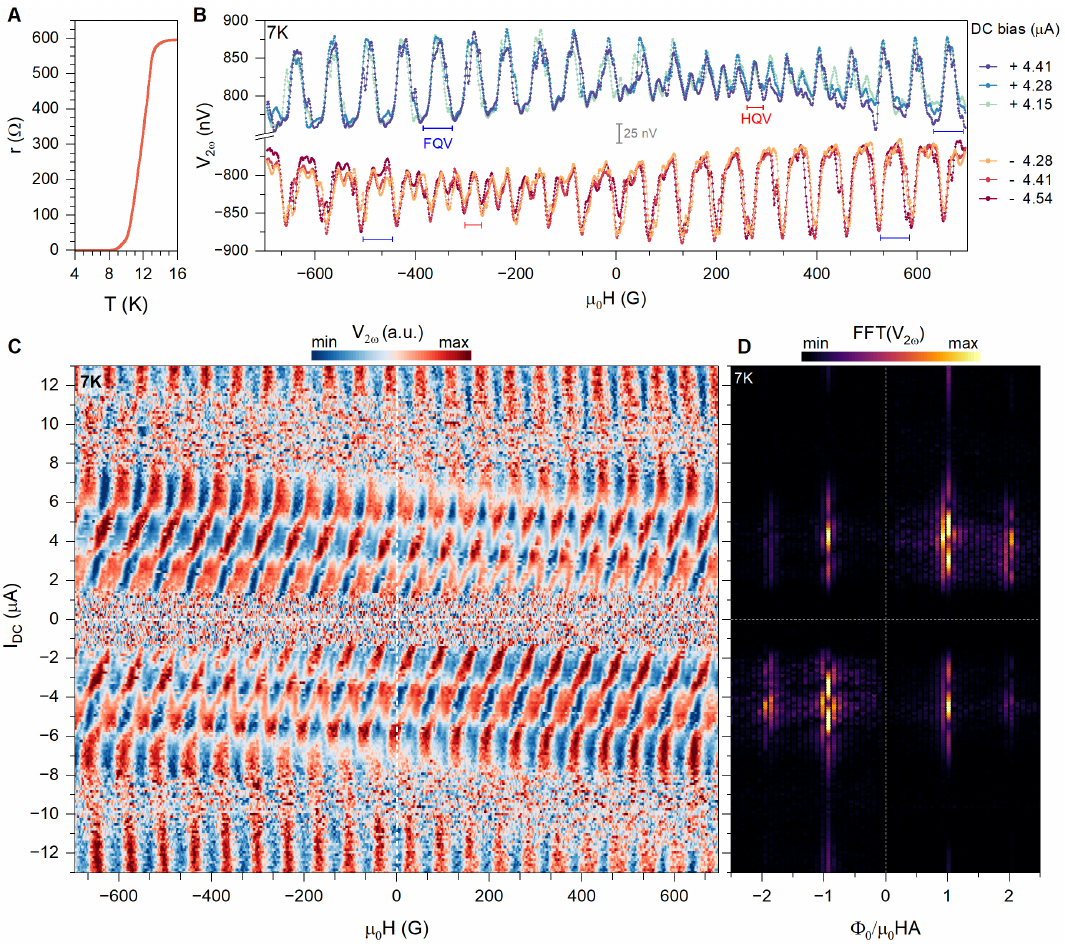}
    \caption{\textbf{Transport and quantum oscillations.} (\textbf{A}) Differential resistance versus temperature. (\textbf{B}) Quantum oscillations of the second-harmonic signal as a function of the perpendicular ($\hat{z}$) magnetic field. The data is collected at 7 K over a range of positive and negative DC offsets. Full- and half-integer quantum oscillations are indicated by blue and red scales, respectively. (\textbf{C}) Matrix plot of the normalized $V_{2\omega}$ as a function of the magnetic field and DC offset (see fig.~S2 for the details of normalization). (\textbf{D}) Fast-Fourier transform of the matrix plot calculated individually over each quadrant highlighting the presence of half-quantum oscillations in the first ($H>0$, $I_{DC}>0$) and third ($H<0$, $I_{DC}<0$) quadrants.}
    \label{fig_mr7K}
    \end{figure}

Fig.~\ref{fig_mr7K}B presents second-harmonic signal as a function of the perpendicular ($\hat{z}$) magnetic field for a range of DC bias offsets collected at 7 K (temperature fluctuations $<$10 mK). At positive DC currents, the signal exhibits consistent oscillations with a period of 67-73 Gauss in both negative (-700 to 0 G) and high positive (400 to 700 G) fields, indicating a full-quantum fluxoid state, as expected from the ring geometry. In the intermediate range (0 to 400 G), a smooth crossover occurs, where oscillations with a periodicity of 30-37 Gauss emerge, indicating half-quantum modulations in the superfluid velocity. At negative DC offsets, the second-harmonic signal exhibits similar FQV and HQV features, with a key difference that the crossover range with HQV modulation now appears at the low negative magnetic fields (-400 to 0 G). The appearance of HQVs alongside FQV oscillations establishes their superconducting origin. Notably, despite the low amplitude of the oscillations ($<25$ nV compared to $>$100 nV for FQV), HQVs reliably appear over a range of DC offsets, confirming the stability and reproducibility of these exotic states. The asymmetric dependence of HQVs on the direction of DC bias and magnetic field rules out trivial scenarios, such as the formation of circulating supercurrents around two adjacent rings.

A matrix plot of $V_{2\omega}$ as a function of the perpendicular magnetic field and DC bias offset (at 7 K) is represented in Fig.~\ref{fig_mr7K}C. For clarity, the second-harmonic signal is normalized: at a given DC offset, the background is removed using a polynomial fitting process, and the resulting signal is normalized (fig.~S2). At large DC offsets ($|I_{DC}|>11 \mu A$), the signal exhibits a pattern of FQV oscillations across the entire range of magnetic field (more than 20 cycles). In the first ($H>0$, $I_{DC}>0$) and third ($H<0$, $I_{DC}<0$) quadrants, as DC offset decreases, each FQV merges with the neighboring FQVs through a decay (splitting) process at approximately $|\pm5|\ \mu A$. Similar pattern of splitting and subsequent convergence of FQVs occurs when the DC offset increases from zero to approximately $|\pm3|\ \mu A$. These two focal patterns meet and interpenetrate at around $|\pm4|\ \mu A$, leading to the emergence of a robust singularity in the phase space, where HQV modulates the oscillatory pattern. The robust singularity is evident from the discontinuity (isolation) of red and blue colors in the V-shape region. It is noted that the absence of signal near zero DC offset arises from the device being in the non-resistive phase. 

In the second ($H<0$, $I_{DC}>0$) and fourth ($H>0$, $I_{DC}<0$) quadrants, FQV exhibits an almost instantaneous $\pi$-shift, accompanied by a reversed sawtooth pattern in the oscillations. Although this $\pi$-shift appears periodically at intervals of $\approx2.6\ \mu A$, it does not lead to the formation of a robust singularity in the phase space, which can be verified from the continuity of blue and red colors. As a result, FQV remains the dominant oscillatory pattern throughout the entire quadrant. This is confirmed by the Fast Fourier Transform (FFT) shown in Fig.~\ref{fig_mr7K}D where the FFT is computed separately for each quadrant. While the FQV signal ($\Phi_0/\mu_0HA = 1$) is present in all four quadrants, a strong HQV frequency ($\Phi_0/\mu_0HA = 2$) appears exclusively in the first and third quadrants. Note that the emergence of HQV coincides with a reduction/splitting in the FFT amplitude of the FQV signal. The shadow (small amplitude) HQV frequency in the second and fourth quadrants originates from the fragile singularity in phase space at instantaneous $\pi$-shift intervals, as well as higher harmonics captured by the FFT process.  The first-harmonic signal $V_\omega$, representing differential magnetoresistance, exhibits a similar pattern of FQV splitting accompanied by signatures of HQV oscillations in its spectrum (fig.~S3).

In Fig.~\ref{fig_mr7K}C, the DC-driven recurring $\pi$-shift is present across the entire magnetic field range. At positive DC offsets, these $\pi$-shifts give rise to robust (fragile) singularity when $H>0$ ($H<0$), while an opposite behavior is observed for negative DC offsets. The clear symmetry between the first and third quadrants, on one hand, and the second and fourth quadrants, on the other hand, indicates the presence of a strong spin-orbit coupling in the superconducting state. Specifically, it follows that $V_{2\omega}(I_{DC},H)=- V_{2\omega}(-I_{DC},-H)$ and $V_{\omega}(I_{DC},H)=V_{\omega}(-I_{DC},-H)$ (fig.~S4). In this context, Fig.~\ref{fig_mr7K}C offers a remarkably compact manifestation of spin-polarized topological superconductivity: full- and half-quantum oscillations capture the superconducting phase and its topological character, respectively, and the symmetry associated with field-current quadrants reveals a rich spin-textured ground state, all encoded directly in the transport signal.

\subsection*{Trapped HQV}
\begin{figure}
    \centering
    \includegraphics[width=\textwidth]{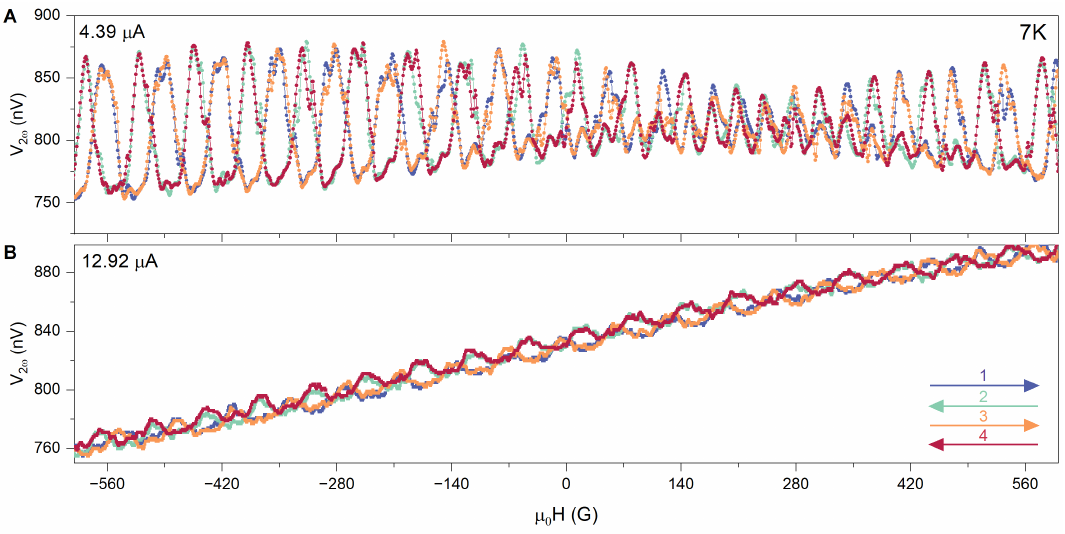}
    \caption{\textbf{Cyclic field sweep and trapped half-quantum fluxoid.} Quantum oscillations of the second-harmonic signal as a function of cyclic sweeps of the perpendicular magnetic field at 7 K for two DC offsets: (\textbf{A}) $4.39\ \mu A$ and (\textbf{B}) $12.92\ \mu A$. The directions of consecutive field sweeps are indicated by arrows in the bottom panel. The quantum oscillations in \textbf{A} are modulated by both FQV and HQV periodicities. The $\pi$-shift between forward and backward sweeps vanishes in the range of 140-420 G where half-quantum vorticity dominates the oscillations.}
    \label{fig_dualsweep} 
\end{figure}

Additional evidence for HQVs emerges from the cyclic magnetic field sweeps. Fig.~\ref{fig_dualsweep} shows second-harmonic oscillations under successive forward and backward sweeps of the perpendicular magnetic field at 7 K, for two DC offsets: $4.39\ \mu A$ where $V_{2\omega}$ is modulated by both FQV and HQV modes, and $12.92\ \mu A$ where only FQV oscillations are present. Although $V_{2\omega}$ follows the same oscillatory pattern during individual forward (blue and orange) and backward (green and red) sweeps of the magnetic field, a clear $\pi$-shift emerges between the two sweep directions in the low-field range. Notably, at $4.39\ \mu A$, the forward and backward sweeps coincide near 200 G, where HQV dominates the oscillations, whereas a relative $\pi$-shift is observed everywhere else. This observation strongly suggests that a trapped half-quantum fluxoid is responsible for the relative $\pi$-shift between forward and backward sweeps. At $12.92\ \mu A$, a similar $\pi$-shift is observed despite the absence of explicit HQV features. Although the oscillatory pattern is dominated by FQV, the presence of the $\pi$-shift points to a trapped half-quantum fluxoid.

\subsection*{Temperature dependence}
In the following, we demonstrate signatures of half-quantum vorticity at zero bias. This is particularly significant, as the emergence of exotic features without external bias underscores the intrinsic nature of topological ground state. Fig.~\ref{fig_temperature}A represents the FFT analysis of second-harmonic spectrum at various temperatures (also see fig.~S5). Here, the Fourier transform is computed over the entire range of the magnetic field (compare the 7 K data to the FFT in Fig.~\ref{fig_mr7K}D). As temperature increases, both FQV and HQV frequencies shift to lower DC offsets due to the narrowing of the non-resistive range. The HQV oscillations emerge at zero bias at $T_c=8$~K, where the spectrum exhibits an oscillatory pattern across the entire range of DC offsets. The corresponding matrix plot of $V_{2\omega}$ at 8 K is presented in Fig.~\ref{fig_temperature}B. At zero DC current, the modulation of half-quantum oscillations in the low-field range is evident through the splitting of individual FQVs, as highlighted by black lines. At higher DC offsets, a recurring $\pi$-shift with a fragile singularity governs the oscillatory pattern at $\approx2.6\ \mu A$ intervals, resulting in the periodic appearance of HQV frequency in the corresponding FFT.

We identify two distinct sets of quantum oscillations, separated by a region where the oscillatory pattern is absent or nearly undetectable. This is evident at 7.5 K, where two strong FFT amplitudes emerge at $\pm 3.5\ \mu A$ and $\pm 11\ \mu A$. While the former aligns with the main oscillatory pattern at 7 K (white dashed lines), extrapolation of the latter (green dashed lines) extends to DC offsets beyond $13\ \mu A$ at the same temperature. The DC offset associated with the emergence of both sets decreases almost linearly with increasing temperature. Within the bias window shown in Fig.~\ref{fig_temperature}A, the oscillatory spectrum is governed by the first one at $T<7$ K, and by the second one at $T>8.5$ K, while in the intermediate temperature range both sets contribute to the oscillatory pattern. Specifically, at 8 K (Fig.~\ref{fig_temperature}B), the two sets of quantum oscillations are roughly separated at $\pm 5\ \mu A$, where the slope of oscillatory pattern changes abruptly, suggesting the presence of two Fermi surfaces with opposite spin texture and slightly different transition temperature. The first-harmonic signal exhibits similar pattern of two separate sets of quantum oscillations (fig.~S6).

\subsection*{Discussion and outlook}
The observation of half-integer periodic oscillations, together with clear indications of a trapped half-quantum fluxoid, offer strong support for the existence of HQVs. Notably, the splitting of FQVs is consistent with the theoretical prediction that a full-quantum vortex may split/decay into two half-quantum vortices in the presence of a driving force~\cite{sigrist_1991,kee_2000,chung_2007}. High-quality perforated superconducting films are known to exhibit signatures of fractional flux, where a flux quantum is shared among multiple adjacent rings~\cite{tinkham_1983,pannetier_1984}. However, these fractional features are symmetric with respect to the applied magnetic field, and do not evolve or split with external parameters such as an in-plane magnetic field or DC offset~\cite{vakaryuk_2011}.

\begin{figure}
    \centering
    \includegraphics[width=\textwidth]{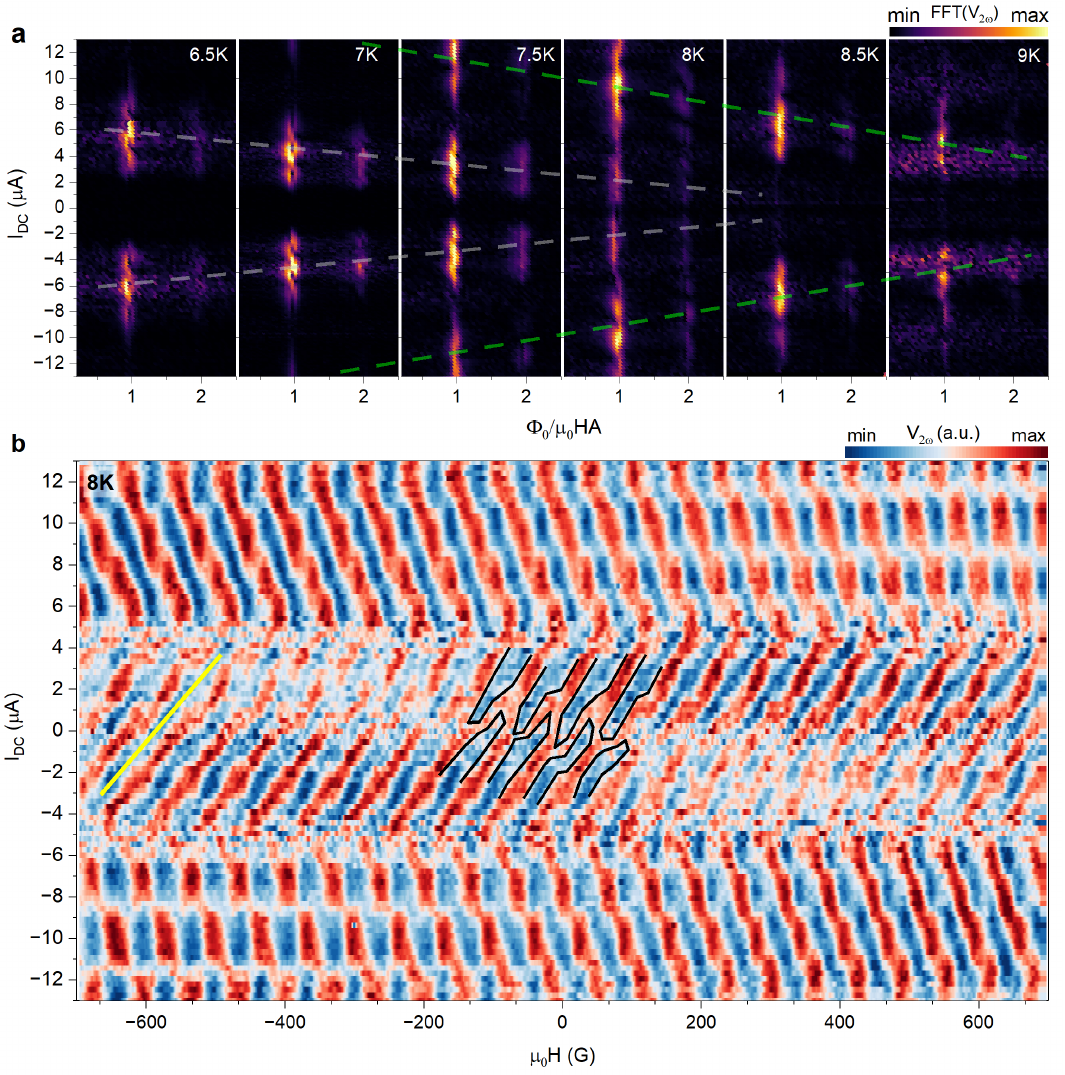}
    \caption{\textbf{Temperature dependence.} (\textbf{A}) Fast Fourier Transform of the normalized $V_{2\omega}$ spectrum at various temperatures (also see fig.~S5). At a given DC offset, FFT is computed over the entire range of the magnetic field. (\textbf{B}) Normalized second-harmonic signal as a function of perpendicular magnetic field and DC offset at 8 K. The white and green dashed lines in \textbf{A}, as well as black lines in \textbf{B} serve as visual guides. The yellow line in bottom panel is used to estimate the effective Zeeman field induced by DC bias current.}
    \label{fig_temperature} 
\end{figure}

A stable HQV is expected to generate an effective Zeeman field ($H_\text{eff}$) due to the velocity mismatch between spin-up and spin-down components~\cite{vakaryuk_2009}. In cantilever torque magnetometry on strontium ruthenate, an in-plane magnetic field was found to be essential for the appearance of half-height magnetization steps~\cite{jang_2011}. While the origin of this coupling is not fully understood, the DC bias current appears to play an analogous role in our experiment. In addition to acting as a tuning parameter to probe the superconductivity transition region---similar in effect to temperature---the DC bias offset gives rise to an effective Zeeman field with a non-zero component along the out-of-plane direction. We determine $H_\text{eff}$ from the slope of oscillatory pattern (yellow line in Fig.~\ref{fig_temperature}B) to be approximately 26 G/$\mu A$. This effective field, when combined with the external magnetic field, gives rise to the recurring $\pi$-shifts at $2.6\ \mu A$ intervals, to fulfill the flux quantization of 70 Gauss. Importantly, the direction of the effective Zeeman field is set by the magnetic field applied during the field-cooling process. When the sample is cooled down in an in-plane ($+\hat{y}$) magnetic field, the effective and external fields are antiparallel in the first and third quadrants, and parallel in the second and fourth quadrants, as can be inferred from Fig.~\ref{fig_temperature}B. In contrast, field-cooling under an out-of-plane ($+\hat{z}$) magnetic field results in parallel (antiparallel) alignment of the effective and external fields in the first (second) quadrant (fig.~S7). These patterns suggest that the field-cooling direction can influence the underlying $\mathbf{d}$-vector texture in the superconducting state. Notably, when the sample is cooled under $-\hat{y}$ field, the HQVs appear symmetrically distributed around zero perpendicular magnetic field in the oscillatory pattern, consistent with an equal spin pairing axis~\cite{leggett_1975}.

Finally, for the thermodynamic stability of HQVs, the spin ($\rho_{sp}$) and ordinary ($\rho_{s}$) superfluid densities must fulfill $\rho_{sp}/\rho_{s}\!<\!(1+\beta)^{-1}$ where $\beta = lw/2\lambda^2$, with $\lambda$ denoting the London penetration depth, and $l$ ($w$) is the side length (wall width) of the ring~\cite{chung_2007}. In this context, reducing the "effective volume" ($lw/2\lambda^2$) through which the spin current flows can promote the stability of half-quantum vorticity. We have fabricated several multiply-connected superconducting devices with varying ring sizes, and found similar transport characteristics for devices with effective ring volume $lw<0.2\ \mu m^2$ (fig.~S8). Our experiments shows that the effective Zeeman field induced by the DC bias current is inversely related to the effective ring volume  $H_\text{eff}\sim(lw)^{-1.12\pm0.11}$, consistent with the expected condition for the stability of HQVs.

We demonstrate the experimental realization of half-quantum vorticity by probing the nonlinear magnetoresistance response in multiply-connected superconducting rings. Direct transport measurements of half-integer quantum states and clear signatures of trapped half-quantum fluxoids provide compelling evidence for the realization of HQVs in solid-state superconducting systems. These findings further establish Fe(Te,Se) as a tunable platform for exploring spin-textured superconducting states. Given the shared electronic and topological properties across the iron-based superconductor family, we anticipate that similar transport characteristics could emerge in related compounds, expanding the landscape for investigating emergent vortex states. A critical next step will focus on probing the vortex core in samples with reduced effective volume to provide conclusive evidence for Majorana zero modes. In parallel, phase-sensitive interferometry may enable detection of braiding and fusion associated with half-quantum vortices, offering a pathway to reveal their non-Abelian statistics. Together, these results lay the foundation for non-Abelian excitations in scalable quantum materials, advancing the pursuit of topologically protected qubits.


\newpage

\clearpage
\bibliography{science_template} 
\bibliographystyle{sciencemag}

%
%
%
%
%
%


\section*{Acknowledgments}
\paragraph*{Funding:}
The work at Brookhaven National Laboratory was supported by the U.S. Department of Energy, Office of Basic Energy Sciences, contract no.~DOE-sc0012704.
\paragraph*{Author contributions:}
M.J.B. designed the experiments, fabricated the devices, performed the measurements, and analyzed the data. M.J.B. and J.U.L. contributed to data interpretation and discussion. Q. L. and G.G. provided the Fe(Te,Se) crystals. K.W. and T.T. supplied the hBN crystals. M.J.B. wrote the manuscript with input form all authors. J.U.L. supervised the project.
\paragraph*{Competing interests:}
The authors declare no competing interests.
\paragraph*{Data and materials availability:}
All data presented in the main text and supplementary materials, along with the corresponding Python code, will be made available in a suitable non-profit online data repository~\cite{dataset}.


\subsection*{Supplementary materials}
Materials and Methods\\
Supplementary Text\\
Figs. S1 to S8\\
Tables S1\\
References \textit{(40-\arabic{enumiv})}\\ 


\newpage


\renewcommand{\thefigure}{S\arabic{figure}}
\renewcommand{\thetable}{S\arabic{table}}
\renewcommand{\theequation}{S\arabic{equation}}
\renewcommand{\thepage}{S\arabic{page}}
\setcounter{figure}{0}
\setcounter{table}{0}
\setcounter{equation}{0}
\setcounter{page}{1} 


\begin{center}
\section*{Supplementary Materials for\\ \scititle}

Mohammad~Javadi~Balakan$^{1\ast}$,
Genda~Gu$^{2}$,
Qiang~Li$^{2,3}$,\\
Kenji~Watanabe$^{4}$,
Takashi~Taniguchi$^{4}$,
Ji~Ung~Lee$^{1\dagger}$\\
\small$^\ast$Corresponding author. Email: mjavadibalakan@albany.edu\\
\small$^\dagger$jlee1@albany.edu.

\end{center}

\subsubsection*{This PDF file includes:}
Materials and Methods\\
Supplementary Text\\
Figures S1 to S8\\
Tables S1

\newpage


\subsection*{Materials and Methods}
Bulk Fe(Te$_{0.55}$,Se$_{0.45}$) crystals were mechanically exfoliated using Nitto blue tape and transferred onto a 300 nm SiO$_2$ substrate inside an Inert nitrogen glovebox (O$_{2}\!<\!.1$  ppm, moisture $\!<\! 0.5$ ppm). The crystal was then fully covered with a thin hBN layer by the dry-transfer method. Since Fe(Te,Se) thin crystals are highly air-sensitive, hBN coverage is essential to prevent degradation during device fabrication. Next, a PMMA(950)/PMMA(495) co-polymer was spin-coated, and electron-beam lithography was used to define small ($\sim 1\ \mu m^2$) patterns, enabling selective hBN removal via reactive ion etching (CF$_4$/O$_2$:50/10 sccm, 40 W, etching rate = 65 nm/min). A second lithography step was performed on a PMMA(950)/MMA(885) co-polymer for contact metallization. Prior to metal deposition, the sample was immersed in a metal-ion-free developer (AZ300) for 30 seconds to gently etch the topmost Fe(Te,Se) layers at a rate of 2.5 nm/min. Electrical contacts (Au/Ti) were deposited by electron-beam evaporation at a base pressure of 10$^{-8}$ torr. After verification of electrical contacts at room temperature, the sample was mounted onto a ceramic chip-carrier and transferred into a dual-beam focused ion beam (FIB) system for milling. The milling process was executed in two stages. In the coarse milling step, a Ga ion beam (30 kV, 230 pA) was used to mill the hBN/Fe(Te,Se) surrounding the active region of the device and ensuring electrical isolation of the contacts. In the fine milling step, a low-current ion beam (30 kV, 20 pA) was employed to mill a pattern consisting of twelve identical interconnect square rings in the active region. Fig.~\ref{figsi_schem}A represents a schematic of the final device structure. To minimize exposure to air, the sample was immediately transferred from the FIB chamber to the cryostat for subsequent measurements. The characteristic length scales of the fabricated devices are summarized in Table~\ref{tabsi_1}, and a scanning electron micrograph of the device with the smallest rings (D5) is in shown Fig.~\ref{figsi_schem}B.

\subsubsection*{Nonlinear magnetoresistance}
The second-harmonic response to an AC excitation offers insight into the response nonlinearity by capturing the gradient of the fundamental harmonic with respect to an independent variable. The oscillatory behavior of magnetoresistance near transition temperature arises from contributions of both Little-Park~\cite{littlepark_1962,littlepark_1964} and vortex-crossing~\cite{sochnikov_2010,sochnikov_2010prb,berdiyorov_2012} effects. The former originates from the periodic modulation of superfluid velocity (T$_c$ oscillations), where the magnetoresistance is given by $R=C(n-\Phi/\Phi_0)^2$. Here, $n$ is an integer number, $\Phi$ is the external flux, and $C$ is a constant. At a fixed AC excitation current, the second-harmonic response is porportional to $V_{2\omega}\propto (n-\Phi/\Phi_0)$, that is (up to a constant) the superfluid velocity, which changes sign at integer (half-integer) intervals leading to FQV (HQV) oscillations. The vortex-crossing effect, on the other hand, is associated with the fluxoid transition induced by the entry and exit of thermally (or DC current) excited vortices into the ring. In this case, the magnetoresistance is given by $R=C\ I^{-2}_{0}[\Delta F(\Phi)/2k_BT]$. Here, $I_0$ is zero-order modified Bessel function of the first kind, $\Delta F(\Phi)$ is an effective potential (free energy) barrier, and $k_BT$ is the thermal energy. Upon a constant AC excitation, the first-harmonic can be approximated by $V_\omega \propto 1-(\Delta F(\Phi)/2k_BT)^2/2$ which leads to $V_{2\omega} \propto -\Delta F(\Phi)/2k_BT$. In this context, the second-harmonic signal in a magnetoresistance experiment is proportional to the changes in free energy of the superconducting condensate. While both mechanisms contribute to the oscillatory magnetoresistance, in the presence of a DC bias current, vortex motion across the ring wall (perpendicular to the DC current) is expected to be the dominant mechanism of dissipation~\cite{bulaevskii_2011, berdiyorov_2012}.

\subsubsection*{Transport}
\begin{figure}
    \centering
    \includegraphics[width=\textwidth]{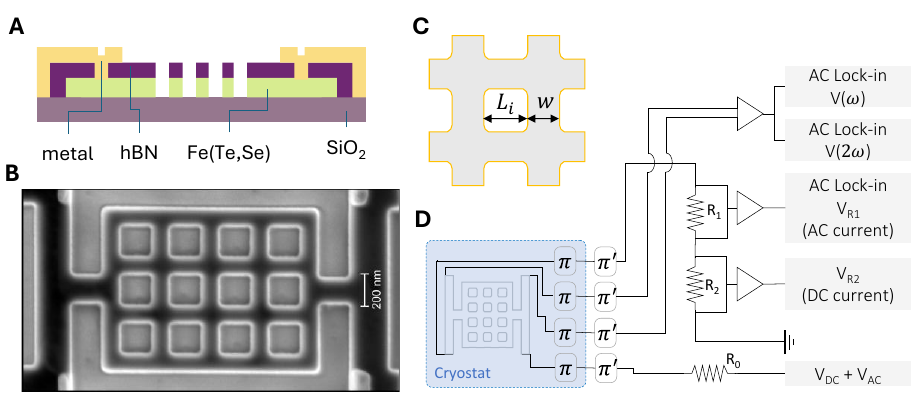}
    \caption{\textbf{Device structure and measurement circuit.} (\textbf{A}) A schematic representing side structure of the device. (\textbf{B}) Scanning electron micrograph of device 5 with the smallest ring size. (\textbf{C}) Schematics representing characteristic length scales of the rings. (\textbf{D}) Measurement circuit. $R_0 = 50\ k\Omega$ is used to source current and $R_{1(2)}$ are reference resistors to measure AC and DC currents. All lock-in amplifiers are synchronized at 23 Hz. Input impedance for AC amplifiers is $100\ M\Omega$. The signal is filtered through double RC ($\pi$) filters with cutoff frequencies at $\pi\!=\!200$ kHz and $\pi'\!=\!2$ MHz. AC voltage drop across the sample was measured simultaneously at the first ($V_{\omega}$) and second ($V_{2\omega}$) harmonics. 
    }
    \label{figsi_schem} 
\end{figure}

Electrical measurements were conducted in a closed-cycle cryostat with a base temperature of 4 K. The system was equipped with an electromagnet mounted on a rotating stage, allowing magnetic fields up to 1 Tesla. Four-terminal lock-in AC+DC method was used to measure voltage drop across the sample under a constant AC excitation of $1\ \mu A$ at $23\ Hz$. A schematic of the measurement circuit is shown in Fig.~\ref{figsi_schem}D. The two-terminal resistance of electrical contacts was generally less than $200\ \Omega$.




\newpage
\begin{figure}
    \centering
    \includegraphics[width=\textwidth]{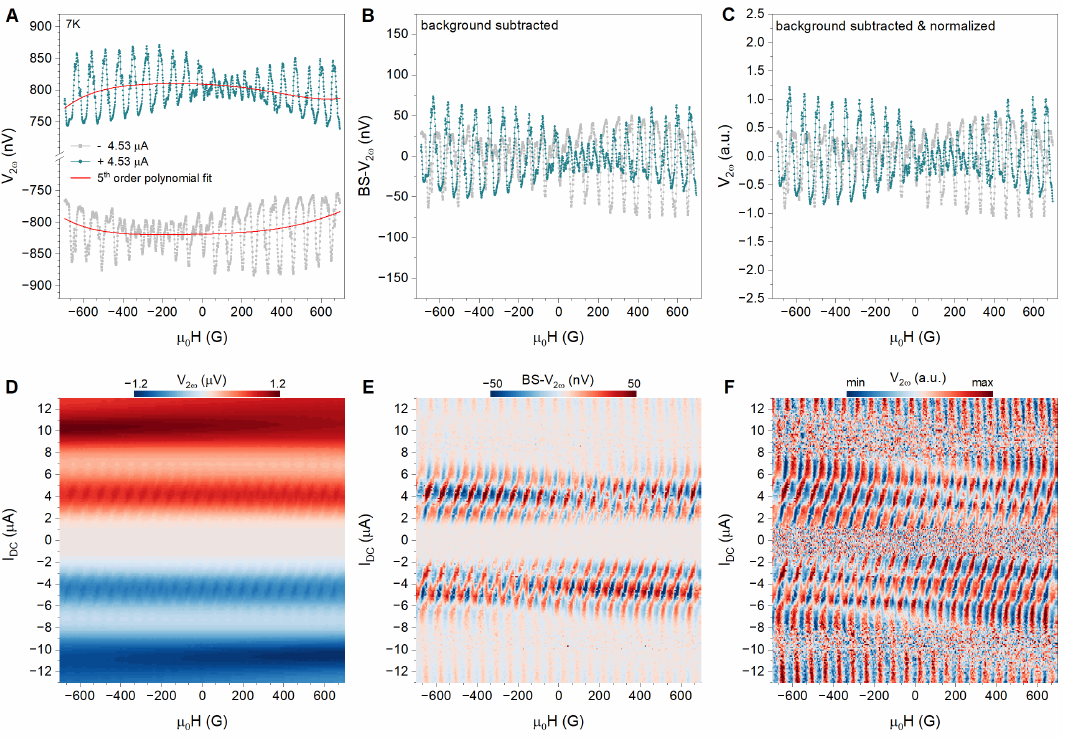}
    \caption{\textbf{Normalization process of the second-harmonic signal.} (\textbf{A}) Raw data as a function of the magnetic field for two DC offsets at 7 K. The red lines represent 5th-order polynomials fitted to the raw data. (\textbf{B}) Background-subtracted $V_{2\omega}$ obtained from subtracting the fitted curves from the raw data. (\textbf{C}) Normalized second-harmonic signal obtained from the background-removed data. (\textbf{D-F}) The corresponding matrix plots of raw data, background-subtracted, and normalized second-harmonic signal as function of the magnetic field and DC bias offset, respectively.}
    \label{figsi_norm} 
\end{figure}
\begin{figure}
    \centering
    \includegraphics[width=\textwidth]{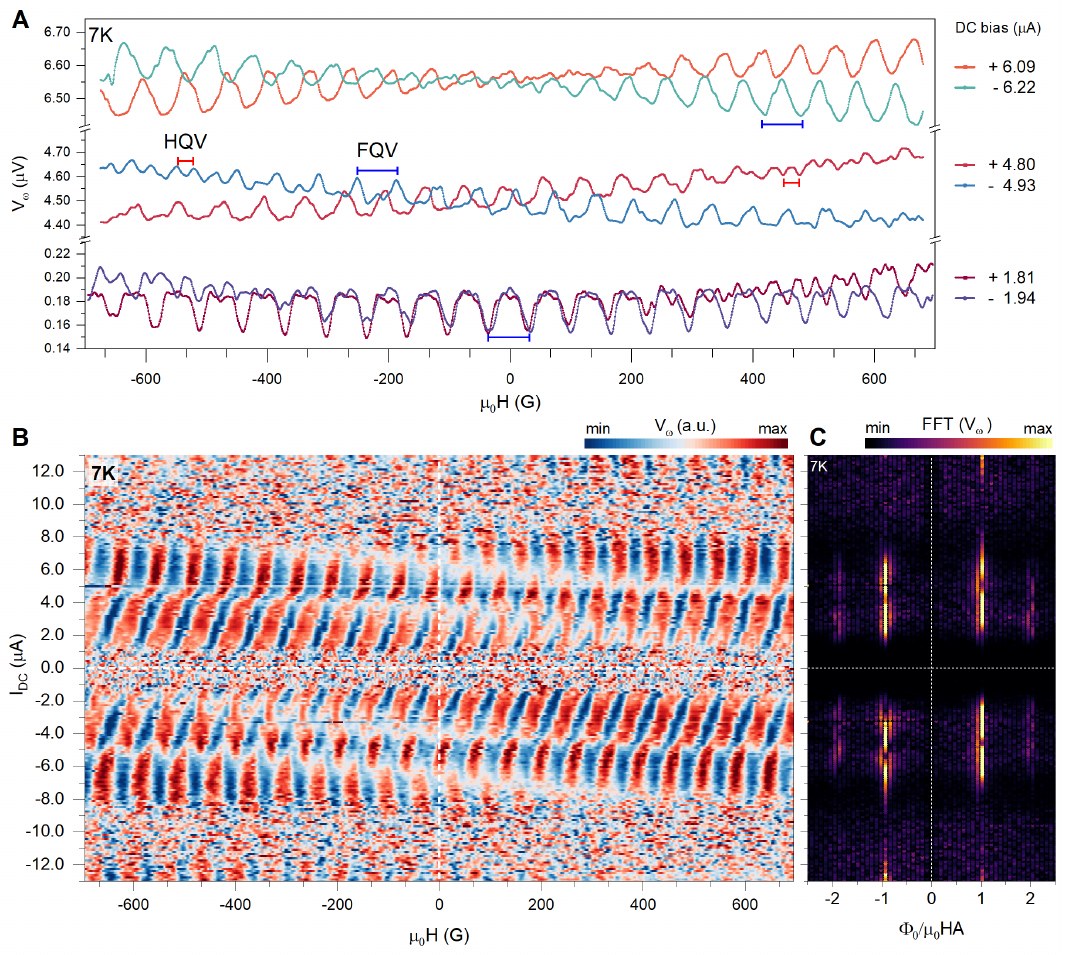}
    \caption{\textbf{Quantum oscillations of the first-harmonic signal (magnetoresistance).} (\textbf{A}) $V_\omega$ as a function of the perpendicular magnetic field for a range of DC offsets. Full- and half-quantum oscillations are indicated by blue and red scales, respectively. (\textbf{B}) Matrix plot of the normalized $V_\omega$ as a function of the magnetic field and DC bias current at 7 K. First- and second-harmonic signals were collected at the same time and identical normalization procedure is used for both signals. (\textbf{C}) Fourier transform of the normalized first-harmonic signal computed individually for each quadrant. The transport characteristics of $V_\omega$ and $V_{2\omega}$ spectra are similar, featuring robust singularity in the first ($H>0$, $I_{DC}>0$) and third ($H<0$, $I_{DC}<0$) quadrants with HQV oscillations, and fragile singularity (spontaneous $\pi$-shift) in the second ($H<0$, $I_{DC}>0$) and fourth ($H>0$, $I_{DC}<0$) quadrants with FQV dominance.} 
    \label{figsi_firstharm} 
\end{figure}
\begin{figure}
    \centering
    \includegraphics[width=\textwidth]{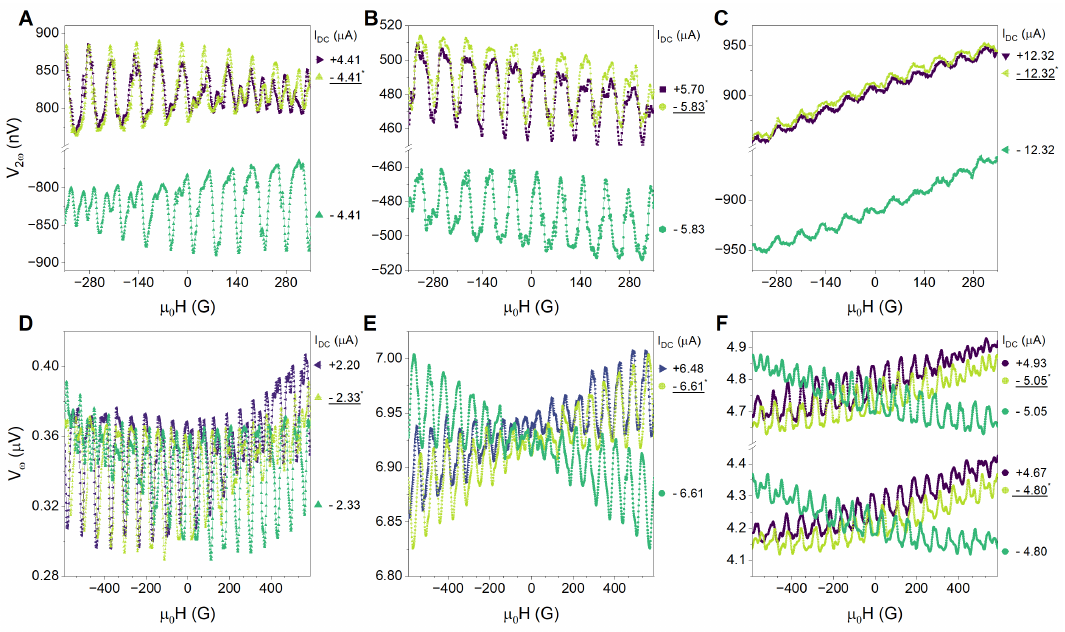}
    \caption{\textbf{Response symmetry.} (\textbf{A-C}) Second-harmonic and (\textbf{D-F}) first-harmonic signal as a function of magnetic field at a range of DC offsets (7 K). The data indicated by star captions ({$^*$}) represent the transformations $V_{2\omega}(-I_{DC},H) \rightarrow -V_{2\omega}(-I_{DC},-H)$ and $V_\omega(-I_{DC},H)\rightarrow V_\omega(-I_{DC},-H)$, shown in the top and bottom panels, respectively. The signal symmetry follows $V_{2\omega}(I_{DC},H)\approx-V_{2\omega}(-I_{DC},-H)$ and $V_\omega(I_{DC},H)\approx V_\omega(-I_{DC},-H)$, suggesting strong spin-orbit coupling in the superconducting state. 
    } 
    \label{figsi_symmetry} 
\end{figure}
\begin{figure}
    \centering
    \includegraphics[width=\textwidth]{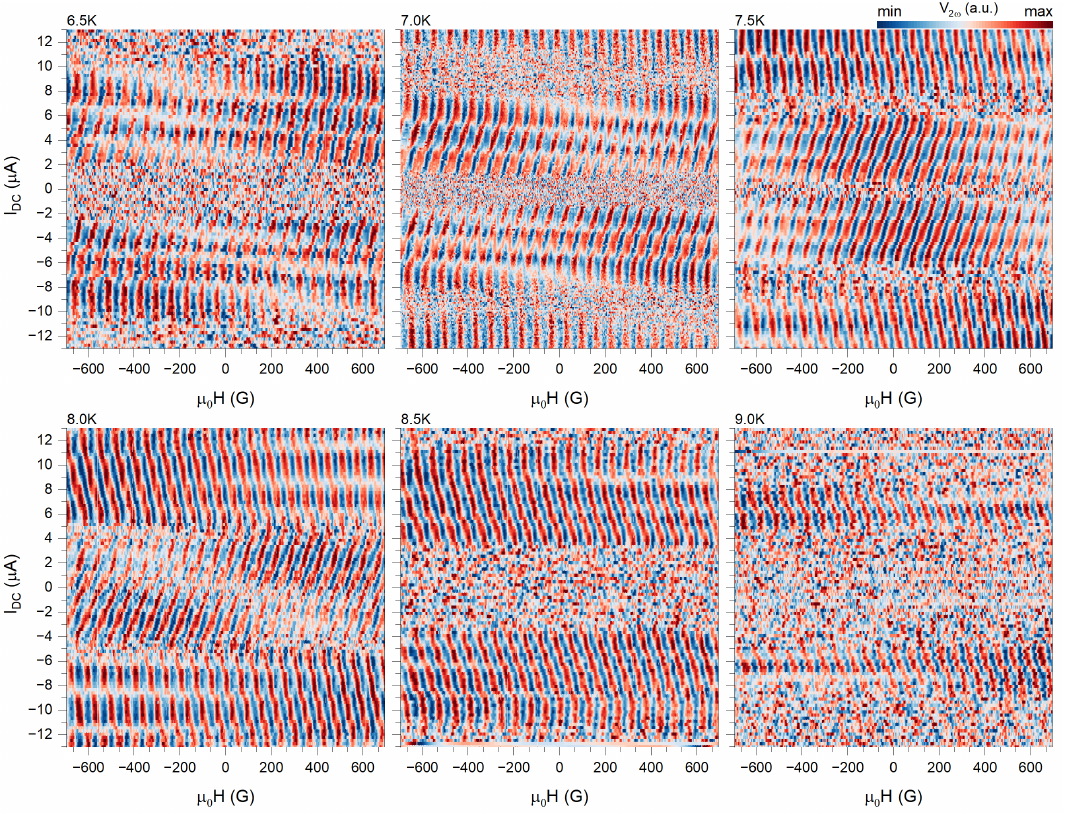}
    \caption{\textbf{Temperature dependence of the second-harmonic response.} Normalized $V_{2\omega}$ as a function of magnetic field and DC bias current at various temperatures. Note that the absence of oscillatory pattern near zero bias at $T<7.5$ K arises from the lack of voltage drop across the sample, while at temperatures $T>8.5$ K is due to the non-oscillatory region between the two separate sets of oscillatory patterns (see main text). Continuous quantum oscillations near zero bias offset at 8 K indicate that the constant AC excitation of $1\ \mu A$ used in the AC+DC magnetoresistance measurements has a minimal effect on the observed oscillatory spectrum.} 
    \label{figsi_V(2w)AllT} 
\end{figure}
\begin{figure}
    \centering
    \includegraphics[width=0.89\textwidth]{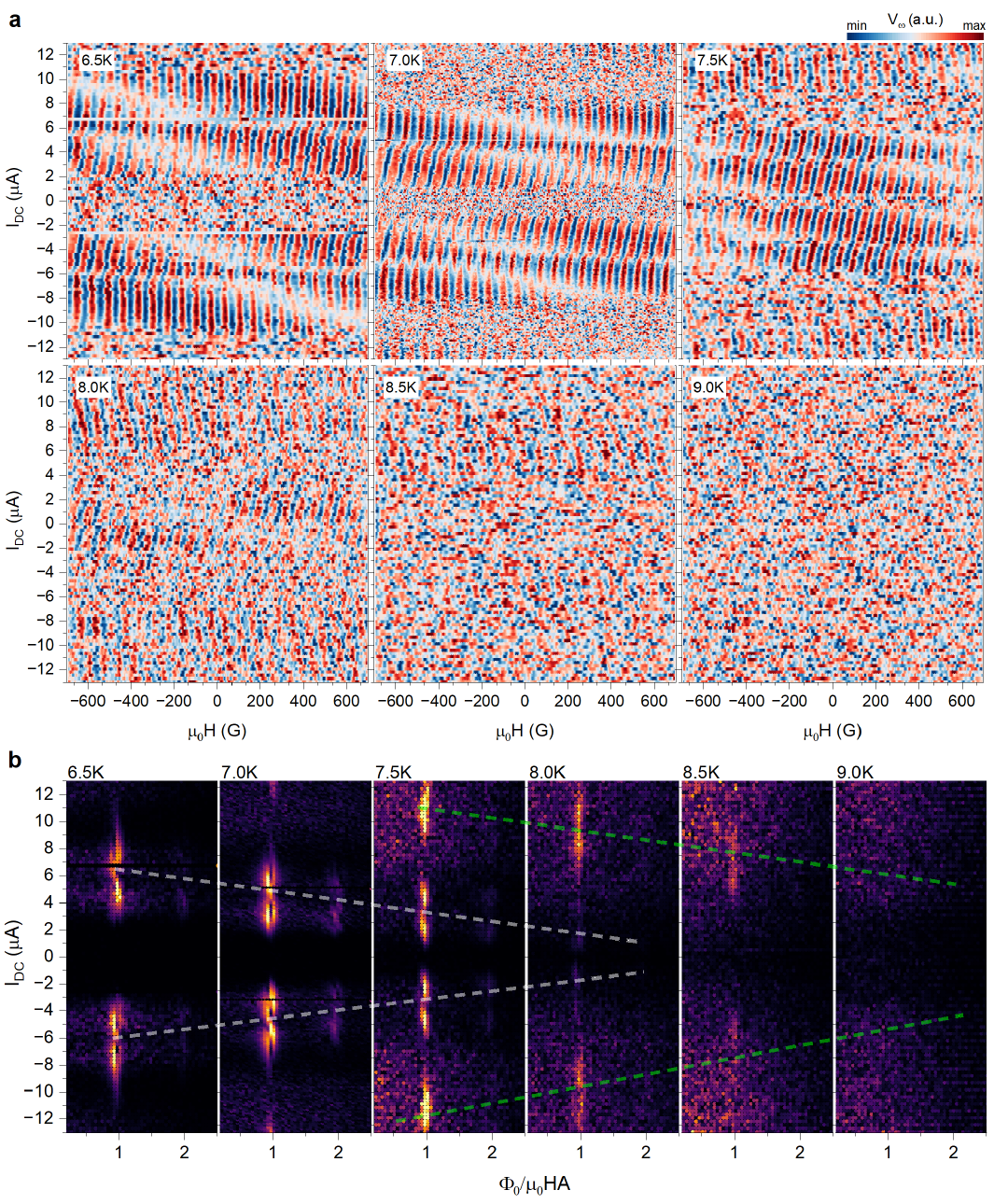}
    \caption{\textbf{Temperature dependence of the first-harmonic response.} (\textbf{A}) Normalized $V_{\omega}$ as a function of magnetic field and DC bias offset at various temperatures. (\textbf{B}) FFT analysis of the first-harmonic spectrum at a given temperature. The FFT is computed across the entire magnetic field. The presence of two distinguishable sets of quantum oscillations in evident as indicated by white and green dashed lines in \textbf{B} (visual guides).} 
    \label{figsi_V(w)AllT} 
\end{figure}
\begin{figure}
    \centering
    \includegraphics[width=0.85\textwidth]{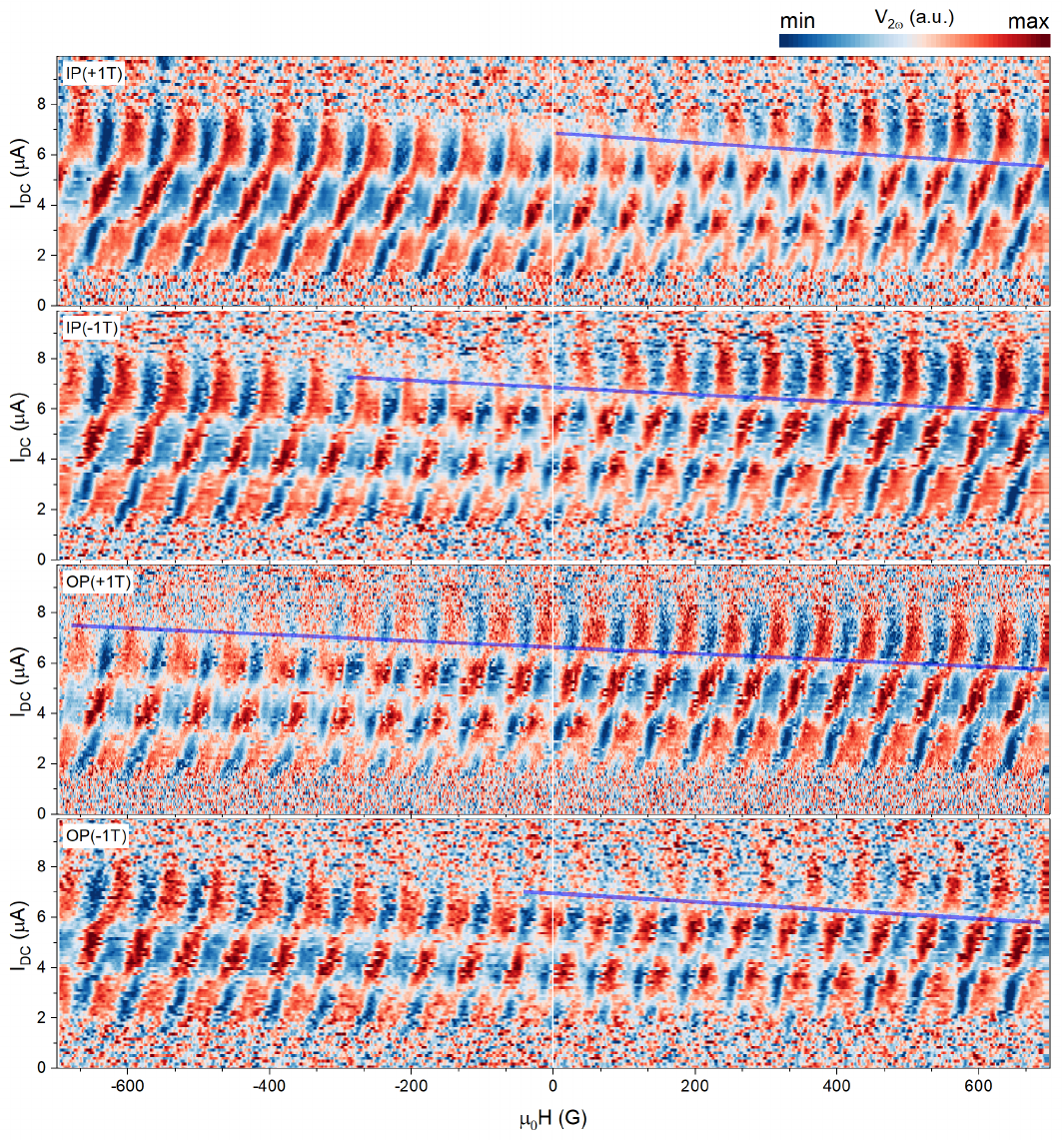}
    \caption{\textbf{Second-harmonic response and field-cooling process.} Quantum oscillations of the normalized second-harmonic signal at 7 K following in-plane (IP, $\hat{y}$) and out-of-plane (OP, $\hat{z}$) field-cooling at $\pm1$ Tesla. The sample was field-cooled from 300 K to 4 K, after which the field was turned-off and the sample warmed to 20 K. The temperature was then set to 7 K, and the measurement was performed. Blue lines are visual guides. For IP (+1 T) and OP (-1 T), the HQV modulation occurs in the first quadrant ($H>0$, $I_{DC}>0$), while for OP (+1 T), the HQV oscillations appear in the second quadrant ($H<0$, $I_{DC}>0$). When the sample is field-cooled under IP (-1 T), the half-quantum oscillations are almost symmetrically distributed around zero field in the oscillatory spectrum.}
    \label{figsi_V(2w)FC} 
\end{figure}
\begin{figure}
    \centering
    \includegraphics[width=\textwidth]{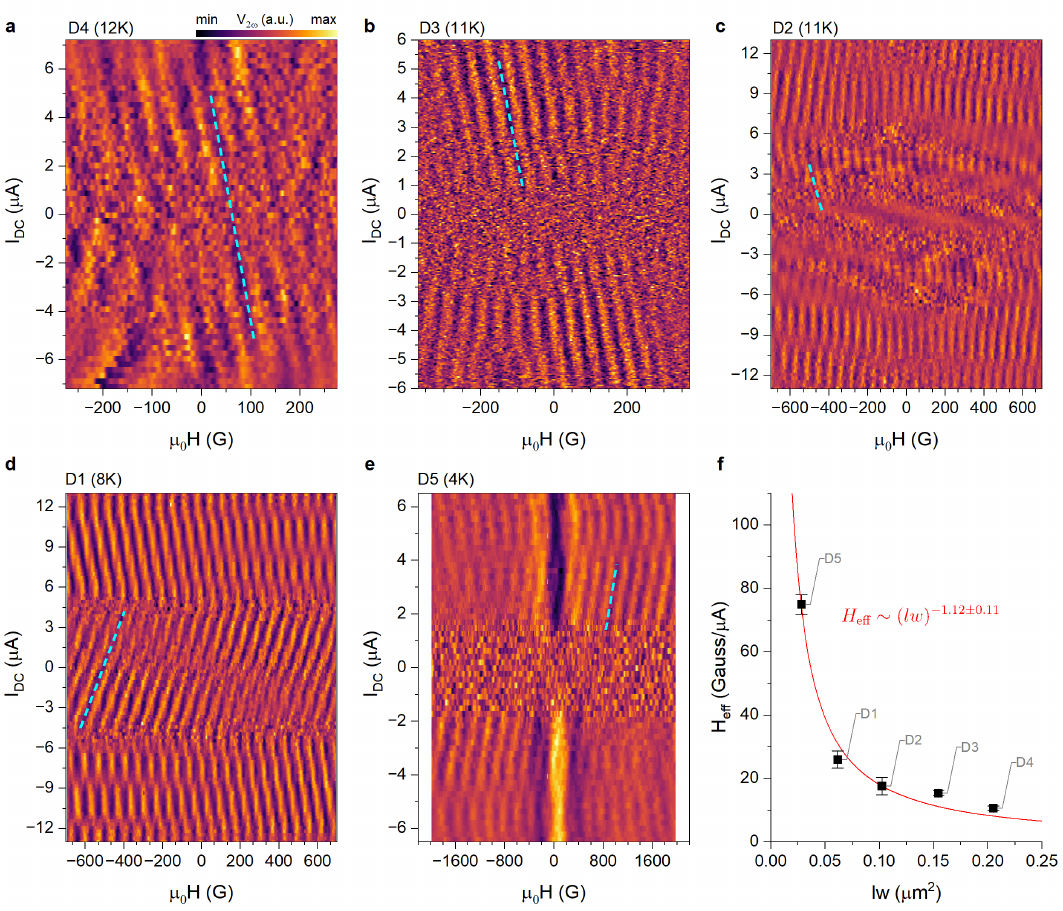}
    \caption{\textbf{Reproducibility and effective Zeeman field.} Normalized second-harmonic response as a function of perpendicular magnetic field and DC offset for five different devices with varying ring size (Table~\ref{tabsi_1}). The data shown in (\textbf{A-D}) is collected at the transition temperature of the corresponding device. For device 5, presented in (\textbf{E}), the data is collected at 4 Kelvin (the lowest available temperature) where $R(4K)/R(20K) = 0.19$. For all devices, the full-quantum periodicity is consistent with the expected flux quantum form the corresponding device geometry (Table~\ref{tabsi_1}). Similar transport characteristics (including FQV splitting, recurring $\pi$-shifts, current-field quadrant symmetry, and DC-induced effective Zeeman field) emerge in devices with effective volume of $lw<0.2\ \mu m^2$. The effective Zeeman field is estimated form the slope of cyan dashed-lines shown in the second-harmonic spectra and is plotted in (\textbf{F}) as a function of the effective volume of the rings, where the red curve represents a power-law fit with an adjusted r-square of 0.98. 
    }
    \label{figsi_Heff}
\end{figure}
\begin{table}
    \centering
    \caption{
    Characteristic length scales of superconducting rings. The thickness of Fe(Te,Se) and hBN crystals are indicated by $t_\text{FTS}$ and $t_\text{hBN}$, respectively. D1 and D2 were fabricated on the same crystal. The effective area of a ring is calculated using $A=l^2(1+(w/l)^2)$, where $l=L_i+w$, and $L_i$ ($w$) denote the inner side length  (wall width) of the ring, determined by averaging over all twelve rings present in the device (Fig.~\ref{figsi_schem}C). The product $lw$ defines the effective volume of the ring. The equivalent magnetic field for a flux quantum is estimated via $\mu_0\Delta H = \Phi_0/A$.}
    \label{tabsi_1}
    
    \begin{tabular}{lccccccr}
        \\
        \hline
         Device & $t_\text{FTS}$ (nm) & $t_\text{hBN}$ (nm) & $L_{i}$ (nm) & $w$ (nm) & $lw$ ($\mu m^2$) & $A$ ($\mu m^2$) & $\mu_0\Delta H$ (G) \\
        \hline
        D1 & 63 & 22 & 406$\pm$19 & 118$\pm$17 & 0.062 & 0.29 & 71.7\\
        D2 & 63 & 22 & 415$\pm$15 & 174$\pm$21 & 0.103 & 0.34 & 54.9\\
        D3 & 86 & 30 &  454$\pm$16 & 227$\pm$15 & 0.155 & 0.52 & 40.1\\
        D4 & 105 & 65 &  401$\pm$12 & 295$\pm$16 & 0.205 & 0.57 & 36.2\\
        D5 & 33 & 27 &  195$\pm$14 & 97$\pm$12 & 0.028 & 0.10 & 217.9\\
        \hline
    \end{tabular} 
\end{table}
\end{document}